# Pedestrian Path Modification Mobile Tool for COVID-19 Social Distancing for Use in Multi-Modal Trip Navigation

Sukru Yaren Gelbal, Mustafa Ridvan Cantas, Bilin Aksun-Guvenc, Levent Guvenc

Automated Driving Lab, Ohio State University

## Abstract

The novel Corona virus pandemic is one of the biggest worldwide problems that occurred within 2019-2020. While death tolls approach 750,000 worldwide and 225,000 just in the US, being mindful of this problem and actively working on precautions is one of the best ways to prevent it from spreading. While hygiene and wearing masks make up a large portion of the currently suggested precautions by the Centers for Disease Control and Prevention (CDC) and World Health Organization (WHO), social distancing is another and arguably the most important precaution that would protect people since the airborne virus is easily transmitted through the air. Social distancing while walking outside, can be more effective, if pedestrians know locations of each other and even better if they know locations of people who are possible carriers. With this information, they can change their routes depending on the people walking nearby or they can stay away from areas that contain or have recently contained crowds. This paper presents a mobile device application that would be a very beneficial tool for social distancing during Coronavirus Disease 2019 (COVID-19). The application works, synced close to real-time, in a networking fashion with all users obtaining their locations and drawing a virtual safety bubble around them. These safety bubbles are used with the constant velocity pedestrian model to predict possible future social distancing violations and warn the user with sound and vibration. Moreover, it takes into account the virus staying airborne for a certain time, hence, creating time-decaying non-safe areas in the past trajectories of the users. The mobile app generates collision free paths for navigating around the undesired locations for the pedestrian mode of transportation when used as part of a multi-modal trip planning app. Results are applicable to other modes of transportation also. Features and the methods used for implementation are discussed in the paper. The application is tested using previously collected real pedestrian walking data in a realistic environment.

## Introduction

The COVID-19 pandemic still continues and the health risk increases every day while the number of infected individuals increases. The pandemic not only produces a health risk, but because of this increase in the number of reported cases, makes individuals more worried about the virus, resulting in a slow-down state for everyday tasks and life in general. Moreover, the significant increase in death tolls worldwide affects the economy of all countries. This number approaches 225,000 just in the US [1]. Precautions such as washing hands [2, 3], wearing masks [4, 5] and social distancing [6, 7, 8] are the easiest to apply and also the most effective ways of preventing the spread of the virus. These precautions are being followed carefully by a wide number of people in both public places, within companies [9] and within universities, etc.

Social distancing is one of the most effective precautions that can be used during the pandemic. However, the scope of social distancing we do in daily life is limited to our own personal space. Although this helps to an extent, since the virus is airborne and can stay in air for a lengthy amount of time [10], the precautionary measure of social distancing would be more beneficial if it can be expanded into a larger scale. This expansion can be done while we walk if we have information of the pedestrians travelling within at least a few miles of distance. As a result, we would be able to know the location of possibly contaminated areas that are caused by individuals who may be carrying the virus. Moreover, it would be possible to determine locations of large gatherings, which would have a very high chance of contamination in a large area.

All of the benefits discussed above are possible if we use our mobile phone to obtain information about pedestrians travelling on streets near us. The information, then, can be used within an application to inform the users about possible contaminated areas, which would result users to re-route or at least be extra careful while travelling through these areas. With the additional features such as automatic rerouting and crowd warnings, re-routing can be handled by the app where optimal non-contaminated pathways are suggested to the user, and user can also be warned about large gatherings around certain areas, which might carry significant amounts of contamination risk. At the end, the app will increase the scope and effectiveness of distancing as a precaution, by providing a beneficial tool for users, which is the main goal of the mobile application developed and discussed in this study.

Additionally, this application can be used within a multimodal transportation environment to minimize the contact with the contaminated areas through all the steps of the travel, walking, driving and parking. This multimodal design is illustrated in Figure 1. Small circles around each module represent different features that would be beneficial for that specific mode of transportation. These features can be benefitted from by implementation in different applications for each mode of transportation or combining them into a single large application similar to Pivot which is a part of Smart Columbus [11]. Smart driving has beneficial features such as distance or fuel efficiency optimized navigation, where smart parking has benefits focused on finding a parking spot that is away from contaminated areas. This is followed by smart walking by the driver and passengers after parking where this study is mainly focused on.

While walking to the vehicle or from vehicle to the destination, this vehicle can also be a bus, metro or ride hailed vehicle if the pedestrian prefers using those. The user can benefit from the app discussed in this study to avoid contaminated paths. Main feature discussed in this study for the app is shown with red circle in Figure 1, which is COVID-19 proximity detection, where the other features such as re-routing and crowd warnings are currently work in progress and are being implemented to obtain a more complete solution for the walking mode of transportation.

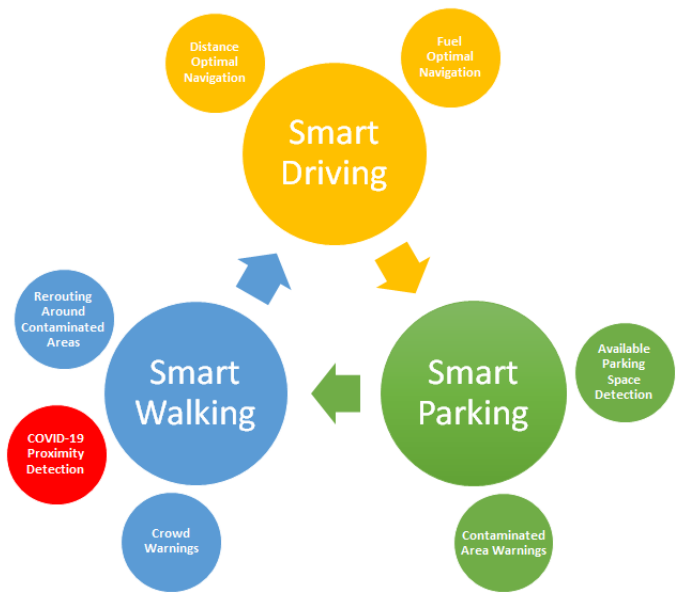

The remainder of the paper is organized as follows. The following Application Data Flow and Features section discusses the designed data flow within the application, followed by determined features to provide the benefits discussed above. The next section, Safe and Possibly Contaminated Area Calculations, discusses the methods used in determining which areas to avoid for the users. This section is followed by the Constant Velocity Model Prediction section, which provides information about how the future location of the user is predicted and what type of warnings are provided. The Implementation and Testing section provides information about the application implementation and demonstrates test results from a few scenarios. The paper ends with the Conclusion and Future Work section, which provides a summary of the work that has been reported within this study and discusses possible improvements that can be realized in the future.

## Application Data Flow and Features

The required data for the application is the location, speed, heading and health information from the mobile phones of users. Location data consists of latitude and longitude, speed and heading are the movement speed and the direction of the pedestrian whereas health data consists of a checkbox that user can mark the user as healthy or unhealthy. It is important to note that there are no health or symptom related testing within the application and the user is free to enter the information as they see fit. It is also important to note that there is no personal data involved in the process.

After the data is obtained, it is sent to other users through a WebSocket server in the cloud. Then, the data is processed within the mobile device to obtain safe and possibly contaminated areas on the map. These areas are then projected onto the map and visualized on the screen for the users, in order to get them informed about any possibly contaminated areas nearby. Moreover, using a simple prediction method, the application predicts the future route and warns user with vibration if the user is moving towards a possibly contaminated area. This vibration feature allows users to receive warnings while the phone is in the pocket, which eliminates the need of looking at the phone screen as an added benefit of using the application. Another significant advantage of vibration based warning system is to be able to inform visually impaired pedestrians as well. While pedestrians will be able to see others on the walkway or walking area ahead and try to execute a socially distant path, this is not possible for visually impaired pedestrians who will need vibration or sound cues for warning. The mobile app presented here is, thus, expected to be much more useful for visually impaired users. Application data flow described above is illustrated in Figure 2.

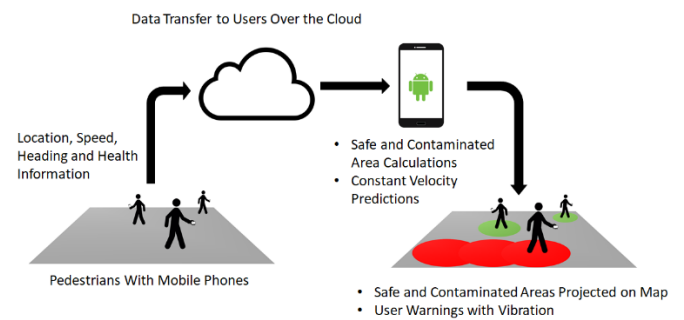

As discussed above, the mobile device (smartphone) application goes through a process starting with obtaining localization data from the phone towards providing useful information and warnings to the other users connected through the Websocket server. The Smart Columbus Operating System (SCOS) of Columbus, Ohio uses a Websocket server which can be used for this connectivity between smartphone users, for example [12]. Application features can be listed as:

- Calculation of safe and contaminated areas according to pedestrian locations and movements.
- Projection of these areas on top of the map, to inform the user about any possibly contaminated areas nearby.
- Prediction of possible intersection of the user with a possibly contaminated area.
- Vibration and visual based warnings to the user in case any future intersection is predicted.

These features for the application are determined in order to implement and test a proof-of-concept application that is beneficial for social distancing, within the scope of this study. They can be improved, or additional features can be added with future work.

## Safe and Possibly Contaminated Area Calculations

The application receives information about the locations of the users, and if the user is healthy or not. Depending on this information, a circular area is created around each user as a safety bubble. In this study, these safety bubbles are measured very conservatively, with a very large safety margin, in order to take into account both the environmental effects such as wind, and the low accuracy on the mobile Global Positioning Sensor (GPS). The standard minimum social distancing amount recommended by CDC is 6 feet [6] and for WHO, this distance is 1 m [7]. But they both suggest: the farther the better. We set the radius of these bubbles as 5 times the suggested value by CDC, which corresponds to 30 feet. This value can be adjusted in the future depending on the research on social distancing for real world use. Figure 3 shows a section of the application screen, which shows how pedestrians and these bubbles are represented on the map. Green bubbles are created for healthy pedestrians and red bubbles are created for possibly unhealthy pedestrians.

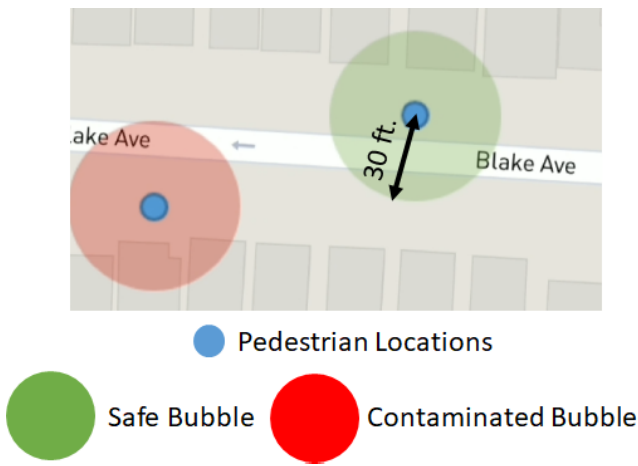

These bubbles are updated in real-time with each GPS update, which is by default 1 Hz in Android phones. Therefore, users can check the screen to see any nearby possibly unhealthy pedestrians, anytime they would like to receive information.

Since the virus can stay in the air for extended amount of time [13], information about the momentary locations and the respective bubbles are not quite enough to capture all the possible contamination information. Pedestrians also contaminate the air around the area they previously walked. Therefore, an additional calculation of previously contaminated areas is necessary. This calculation was designed as a decay based algorithm within the app. The locations pedestrians are walking through at each update are kept in memory and a decay function is applied on these areas to simulate the slow dissipation of the possible contamination. This decay function is initially implemented as a linear decay of the contamination percent with respect to time, as follows.

$$C(t) = C_{start} - \frac{100}{C_{start}} \frac{t}{t_{airborne}} \quad (1)$$

where $C$ is the contamination percentage, $C_{start}$ is the initial value of the percentage contamination that the corresponding pedestrian causes while walking through the location, $t_{airborne}$ is the maximum time value that the virus can stay airborne and $t$ is the time passed after the area was contaminated. $C_{start}$ and $t_{airborne}$ are designed as adjustable parameters for the app, which can be changed depending on future virus related research or depending on the additional information such as whether the user is wearing a mask or not. After the contamination value reaches zero, the contaminated area is freed. More complex formulations are researched and investigated by other researchers in physics and fluid dynamics fields [10, 14]. The formula used in current implementation can be changed later depending on future research in the area.

New values are calculated through the locations the pedestrians are walking, as a result of the equation above. Contaminated areas with reduced percentage are visualized as circles, similar to bubbles around the pedestrians. As the contamination amount decays with time, the circles are visualized with increasing transparency. This results in a decaying trail of contaminated areas behind the pedestrian. The effect can be seen clearly in Figure 4, which shows a section of the app screen with the visualization discussed above.

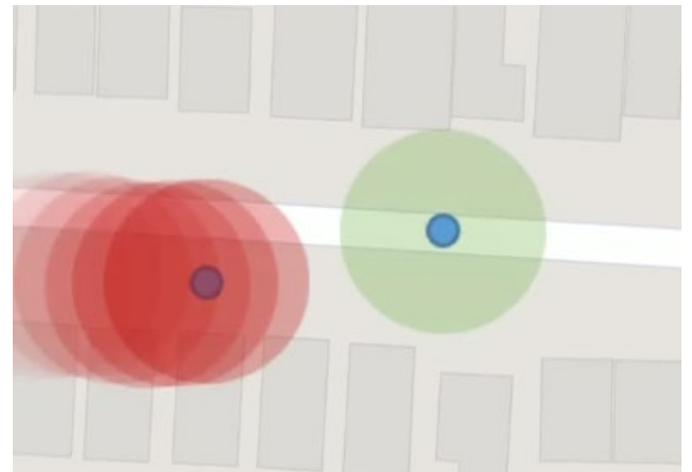

It is important to note that the virus can stay airborne for up to 3 hours [13]. Since 3 hours is very long time for the visualized area to decay completely, the effect created in Figure 4, as well as the figures within the Implementation and Testing section were made possible by significantly reducing the amount of time the virus stays airborne, to about 6 seconds. This reduction was done only for demonstration purposes and is not intended for the actual usage of the app. Also, as seen in the figure, this prolonged contamination and decay effect is applied only for unhealthy pedestrians, which can be depicted from red color. Green bubbles belong to healthy pedestrians, which doesn't cause any contamination related effects while walking.

## Constant Velocity Model Prediction

In order to warn the users for possible intersections with possibly contaminated areas, pedestrian movement should be predicted, depending on the intended time before the area intersection. After determining the contaminated areas and personal bubbles around the pedestrians, future predictions for paths of the users are calculated by a method based on a simple constant velocity model. If we denote pedestrian heading in terms of radians with $\theta$ and speed in terms of m/s with $v$, components of the velocity vector $\vec{v} = [v_x, v_y]$, can be simply calculated as,

$$v_x = v\cos(\theta)$$
$$v_y = v\sin(\theta) \quad (2)$$

If we define location vector in terms of meters as $\vec{u} = [x, y]$, using the calculated velocity vector, future location of the pedestrian can be calculated simply as,

$$\vec{u}_p = \vec{u}_c + \vec{v}_c(t_p - t_c) \quad (3)$$

where $t_p$ denotes the time for the future location to be predicted, and $t_c$ denotes the current time. Location vectors $\vec{u}_p$ and $\vec{u}_c$ represent predicted location and current location of the user respectively. Velocity vector $\vec{v}_c$ represents the current velocity of the user. Note that it is possible to use a Kalman filter or a data driven neural network algorithm to predict future path with more accuracy.

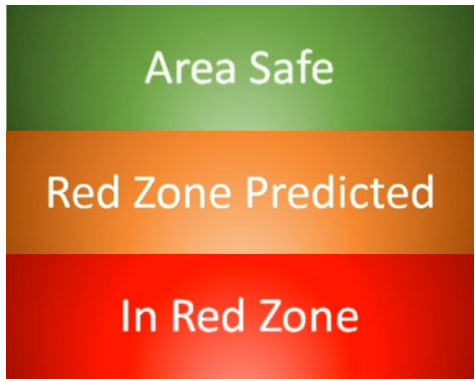

After the calculation of the future location, three different states are possible for the user and are shown in Figure 5. If there is no predicted intersection with any contaminated area, the user receives the "Area Safe" state. If there is any intersection predicted within a certain safety time, which was set as 3 seconds in our case, "Red Zone Predicted" state is issued to the pedestrian with intermittent vibration as a warning. During this state, the pedestrian can check the phone to receive info about the exact location of the contamination to re-route around it, or he/she can just change direction after feeling the vibration. Rerouting the pedestrians optimally is a separate task by itself and is not treated within the scope of this study. However, it is currently being worked on along with providing sound cues to visually impaired users and planned to be implemented in the near future.

Third and final state is the "In Red Zone" state, which indicates that the user's bubble is intersecting with a contaminated area bubble. The user is warned visually as well as with continuous vibration. The user is expected to leave the contaminated area to keep the travel safe.

## Implementation and Testing

An Android application was developed with the implementation of the methods and algorithms discussed in the previous sections. Android Studio was used to develop this as a Java based application, using several modules for obtaining measurements and visualizing the information on the map. The overall implementation diagram for application testing is displayed in Figure 6.

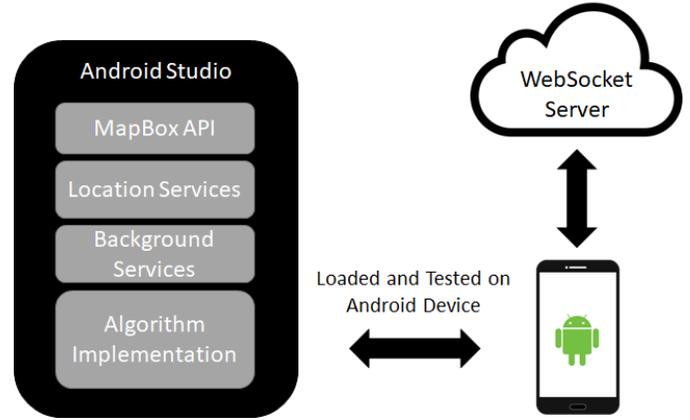

In order to project the locations of the users and the possibly contaminated areas on a map, a free to use Application Programming Interface (API), MapBox API was used within the application. This API allows the application to visualize the information more efficiently for the users, by showing the information on top of a map. Location Services were used to obtain GPS location from the user and Background Services were used to allow the app to run in the background, even when the users are carrying their mobile devices in screen-locked mode in their pocket. The Algorithm Implementation module includes all the calculations discussed in the previous section, in order to determine safe and contaminated areas, as well as to make location predictions for the users.

After the application was developed in Android Studio, it was loaded into Android Device and tested. The test is for the case in which the application is used by multiple travelling users. Two different scenarios were prepared for testing of the app. These scenarios were tested in the setup described above and the screen of the mobile device was recorded as a video recording. Results are provided as screenshots and video links for each scenario under their respective sections below.

### Scenario 1

Scenario 1 is created as a simple test scenario where two pedestrians directly pass through each other's side. The user of the app is healthy (green bubble), but the other pedestrian is unhealthy (red bubble). At start, they both move towards the same direction until a point where the user decides to cross the street and walk towards the other pedestrian. Overall testing process can be seen in Figure 7 with screenshots taken from the application screen and explained in detail below. Direction of motion for the user was marked with a blue arrow since it is not straightforward to recognize just by looking at the screenshots.

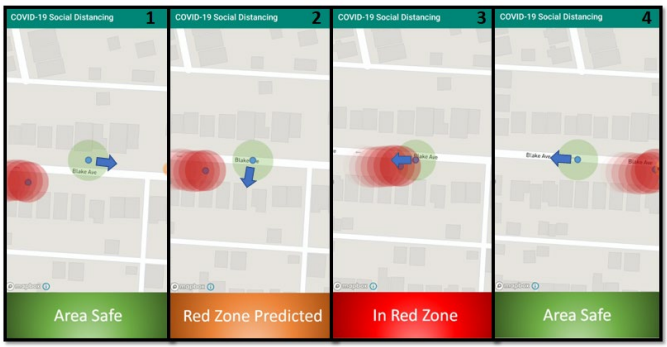

As seen in Figure 7 picture 1, the user starts in a safe area and moves towards a safe direction. But in picture 2, the user turns to the right, crosses the street and moves towards a contaminated area. This results in a prediction of intersection with the contaminated area. The user is given a "Red Zone Predicted" warning accompanied with vibration. The user still continues to walk towards the red area as seen in picture 3, resulting in the user crossing the red zone, next to an unhealthy pedestrian. The user is given an "In Red Zone" warning and also warned with vibration. At the end, the user exits the contaminated area through the other side and returns to "Area Safe" again as seen in picture 4. Note that the virus airborne time is reduced to 6 seconds for this scenario to demonstrate the decay effect on the app screen. The video for this scenario can be accessed through the link given in [15].

*Scenario 2*

The second scenario is slightly more complex as compared to the first one. In this scenario, the user takes the warnings into account and tries to wait until the prolonged contamination effect dissipates until crossing the street to reach his/her destination. The process of testing for this scenario, similar to the first one, can be seen from the screenshots in Figure 8 and is described in detail below. The direction of the movement for the user is shown with a blue arrow in this figure as well, for the same reason as the previous one. Pictures with no blue arrow indicate that the user is not moving.

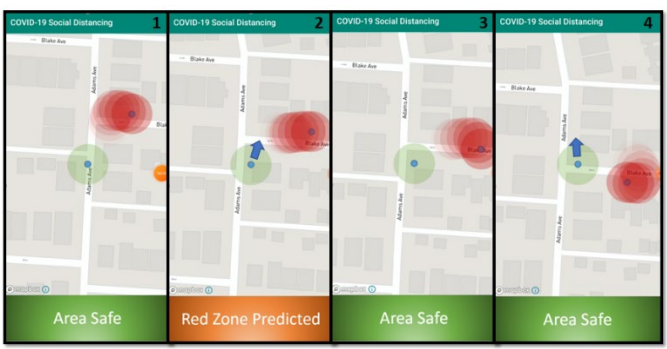

As seen in Figure 8 picture 1, the user starts waiting at a location and plans to go towards North. But there is a contaminated area nearby because of the virus staying airborne. While the user approaches towards this area in picture 2, a warning is issued because of a future intersection with the contaminated area is predicted. After receiving the warning, the user waits for the contamination to dissipate, in picture 3. Finally, after the contaminated area disappears from the planned path, the user crosses the street safely, as seen in picture 4. It is important to note again that the time for the virus to stay airborne was set to 6 seconds for demonstration purposes. The video for this scenario can be accessed through the link in reference [16].

## Conclusions and Future Work

During these times when the Coronavirus pandemic is a significant health problem for most countries in the world, any counter-measure for prevention of the spreading of the virus would immensely help people. Active measures such as social distancing are proven to be highly effective for preventing the spread, along with wearing masks.

This paper presents an Android application that can be used in a multimodal smart mobility environment, in order to aid users achieve better social distancing, with features involving large scale contaminated area information, predictions and warnings. The methodology used in several aspects of the application were discussed, as well as information about the data flow and implementation structure were provided. After the implementation, the application was also tested within a realistic environment, and results for two different example scenarios were shown with screenshots as well as video links. It is important to emphasize that, methods used in this study are all intended for a proof-of-concept scale implementation. Therefore, each feature has room for improvement as future work.

Crowd warnings, large scale heatmap information and optimal routing for avoiding contaminated areas are currently being worked on and planned to be implemented as additional features for the app in the future. Rerouting feature was illustrated on top of a navigation image from the Pivot app [17] in Figure 9, as an example. The illustration depicts a pedestrian who intends to take a bus and, therefore, navigated to a bus station by a multimodal trip planning app. Picture "a" shows the default navigation that aims for the shortest path, whereas picture "b" shows a modified path where a rerouting of the contaminated area showed in red circle exists on the default path. As the future work, this feature combined with other intended features discussed earlier, would make the app an easy to use and very effective tool to reduce the spread of COVID-19.

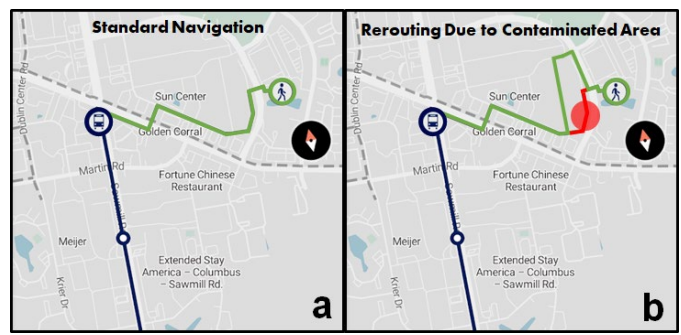

The mobile app and the method presented here can be extended to other applications of conflict zone avoidance. It can be re-purposed with audio cues to help pedestrians with visual impairments and the risk zone definition can then extended to include other possible risks like construction, accident location, problematic side walk etc [18]. The mobile app can also be integrated with on-demand, ride-hailed autonomous vehicles to improve mobility [19]. The mobile app can also be used in path planning of vehicles including autonomous ones for avoiding risk zones automatically and can be integrated with path planning and robust path tracking control [20], [21], [22], [23], [24], [25], [26], [27], [28], [29] [30], [31], [32], [33], [34], [35], [36], [37]. A realistic model-in-the-loop or a hardware-in-the-loop simulation environment can also be used for further development of the mobile app [38], [39], [40], [41], [42], [43], [44], [45], [46], [47], [48] and state feedback control [49] can be used for path tracking along with robust control [50], disturbance observer control [51], model predictive control [52] and other methods.

vehicle (CAV) applications," in *SAE World Congress and Experience*, Detroit, 2019.

## Definitions/Abbreviations

| | |
|---|---|
| **CDC** | Centers for Disease Control and Prevention |
| **WHO** | World Health Organization |
| **COVID-19** | Coronavirus Disease 2019 |
| **GPS** | Global Positioning Sensor |
| **API** | Application Programming Interface |

## Contact Information

Sukru Yaren Gelbal

The Ohio State University

Electrical and Computer Engineering Department

gelbal.1@osu.edu

1320 Kinnear Road, Columbus, OH, 43212